\documentclass[%
  nofootinbib,
  eqsecnum,
  superscriptaddress,
  a4paper,
  showkeys,
  showpacs,
]{revtex4-1}
\usepackage{amsmath}
\usepackage{amssymb}
\usepackage{hyperref}
\usepackage[dvipsnames]{xcolor}
\usepackage{xcolor}

\begin{document}
 \title{Different types of torsion and their effect on the dynamics of fields}
  \author{Subhasish Chakrabarty}
  \email{subhasish.chy@bose.res.in}
  \author{Amitabha Lahiri}
  \email{amitabha@bose.res.in}
  \affiliation{S. N. Bose National Centre for Basic Sciences\\
  Block - JD, Sector - III, Salt Lake, Kolkata - 700106}

\begin{abstract}
{

One of the formalisms that introduces torsion conveniently in gravity is the vierbein-Einstein-Palatini (VEP) formalism. The independent variables are the vierbein (tetrads) and the components of the spin connection. The latter can be eliminated in favor of the tetrads using the equations of motion in the absence of fermions; otherwise there is an effect of torsion on the dynamics of fields. We find that the conformal transformation of off-shell spin connection is not uniquely determined unless additional assumptions are made. One possibility gives rise to Nieh-Yan theory, another one to conformally invariant torsion; a one-parameter family of conformal transformations interpolates between the two. We also find that for dynamically generated torsion the spin connection does not have well defined conformal properties. In particular, it affects fermions and the non-minimally coupled conformal scalar field.

}
\end{abstract}


\maketitle

\section{Introduction}
Conventionally, General Relativity (GR) is formulated purely from a metric point of 
view, in which the connection coefficients are given by the Christoffel symbols 
and torsion is set to zero \emph{a priori}. Nevertheless, it is always interesting to consider a more general theory with non-zero torsion. The first attempt to formulate a theory of gravity that included torsion was made by Cartan~\cite{Cartan1922}. This is also known as Einstein-Cartan theory.
It was then further developed by Kibble~\cite{Kibble:1961ba}, Sciama~\cite{Sciama:1964wt}, and later 
Hehl~\cite{Hehl:1971qi}, relating torsion to the spin angular momentum of matter, in particular 
fermionic matter. Torsion induces an effective self-interaction among fermions thereby 
making the Dirac equation nonlinear, analogous to the Nambu-Jona-Lasinio model~\cite{Nambu:1961tp, Nambu:1961fr}. This feature has been exploited in particle physics, for example
in~\cite{Fabbri:2012zd} it has been suggested that torsion induces interactions among 
leptons identical to the weak leptonic interactions in Weinberg's standard model~\cite{Weinberg:1975gm}. 
The spin-spin interaction induced by minimal coupling of fermions with torsion has been shown to help replace big bang singularity with a cusp-like bounce~\cite{Poplawski:2011jz}. In cosmological models~\cite{Vignolo:2014wva, Myrzakulov:2010vz, Nikiforova:2017saf} the possibility of a self accelerating universe has been considered from a torsion point of view. Also it has been suggested that four-fermion interaction originating from spin-torsion coupling can be seen as dark energy~\cite{Poplawski:2011wj}. The possibility of inflationary phase in early universe has been discussed with spin and torsion without the need of any extra fields~\cite{Akhshabi:2017lyg}.

The torsion-less limit of torsion gravity cannot always be taken continuously~\cite{Fabbri:2014kea,Fabbri:2014dxa}. We are interested in a torsion theory of gravity which in the torsion-free limit reduces to Einstein's GR. One way of introducing the torsion field into the theory as a dynamical variable would be to add it in as the antisymmetric part of the connection coefficients. However, the way to couple torsion to other fields, particularly to fermions, is not obvious in this approach. A more transparent and geometrical way of introducing torsion is to work with the first order Palatini formulation of gravity, using local orthogonal coordinates or frame fields called tetrads or vierbeins, 
and a local Lorentz connection called the spin connection~\cite{Kibble:1961ba, Peldan:1993hi, Hehl:1994ue}. 
We will call this vierbein-Einstein-Palatini (VEP) formalism. When gravity is coupled to only bosonic fields, this formalism reduces to the usual metric formalism of GR
on-shell, where the spin connection can be written in terms of the tetrads and their derivatives. 
If there are fermionic fields contributing to the stress-energy tensor, the spin connection has 
torsion components and remains independent. This formalism is particularly useful for writing a 
Lagrangian for fermionic fields on curved space-time~\cite{Fock:1929vt, Pollock:2010zz,Hehl:1976kj}, 
as it highlights the spin connection as being analogous to a gauge field. In addition, the VEP 
formalism serves as the link between GR and BF theories of gravity~\cite{Freidel:1999rr, 
Capovilla:2001zi, Montesinos:2011ak, Celada:2016jdt}.

In this paper we investigate conformal transformations of vierbeins and spin connections. The 
motivation for this investigation is twofold. First, if the VEP action is exactly 
equivalent to the second-order Einstein-Hilbert action of gravity, all matter fields should couple to 
gravity `in the same way' in both formulations. More precisely, the corresponding stress-energy
tensors for matter should be equivalent in the two formulations, and matter fields ought to 
transform in the same way in both the formulations. While this is a trivial issue for minimally 
coupled matter fields, it turns out that for non-minimally coupled fields, such as the 
conformally coupled scalar which we investigate here, the field equations behave differently under conformal transformations in the two formulations.
The other motivation is to study the conformal transformation of spin connection, which 
is useful in studying the conformal properties of fermions propagating on a curved background. Spin connection has a part that acts as torsion, thus its behavior under conformal transformations affects that of fermions.

We will consider different possibilities of how torsion is affected by conformal transformations. 
First we will discuss Nieh-Yan theory~\cite{Nieh:1981xk}, in which torsion was considered to 
``play the role of gauge transformation for the conformal transformation group." 
However, when torsion is taken to zero, this theory does not reduce to pure Einstein gravity, i.e. GR based on pure Riemannian geometry. Next we discuss a theory where torsion remains invariant under conformal transformations. We show that Nieh-Yan theory and the one with invariant torsion 
correspond to two limits of a general transformation of the spin connection which interpolates between these two limits.
We also consider dynamically generated or on-shell torsion which is the expression for torsion obtained by solving the equation of motion. We show that torsion, dynamically generated by the Dirac field as in~\cite{Hehl:1976kj,Nieh:2007zz}, transforms homogeneously under conformal transformation. In other words, unlike in Nieh-Yan theory, on-shell torsion does not have any inhomogeneous conformal transformation. We also discuss the possibility of on-shell torsion being generated by a conformal scalar field. However, for the scalar field, we find that on-shell torsion indeed transforms inhomogeneously.

Conformal transformations were introduced by Weyl in an attempt to unify electromagnetism 
and general relativity~\cite{Weyl:1919fi}, and have been useful in studying various properties 
of curved spacetimes~\cite{Faraoni:1998qx}. 
Conformal transformations have been widely used in studying asymptotic flatness and initial value problem~\cite{Wald:1984rg, Friedrich:2008tj, Friedrich:2007db, Garfinkle:1986vt, Friedrich:2012iz}, 
propagation of massless fields on a gravitational background~\cite{Sonego_2, Noonan:1995jy,
Paetz:2013nga, Behroozi:2005md, Bracken:1982ny, Barut:1981mt, Gursey:1956zzb, Dirac:1936fq, Perlick:1990df}, exact solutions~\cite{Bergh:1980mta, Van_2, Van_3, Tsamparlis:2015qua, 
Said:2012xt, Verbin:2010tq, Bekenstein:1974sf} and other problems where
scale-independence is fundamental to our understanding of the system. Conformal invariance is 
also important in the study of quantum field theory on curved spacetime~\cite{Birrell:1982ix,
DeWitt:1975ys, Wald:1978pj, Ford:1997hb}. It has been suggested that cosmology based on conformal gravity, or more specifically based on the Weyl tensor, can provide alternatives to the usual cosmologies with dark matter and cosmological constant~\cite{Mannheim:1988dj, Mannheim:1989jh}. 

A conformal transformation is the scaling of the spacetime metric $ g_{\mu\nu} $ with a strictly positive, smooth function $ \Omega^2 $, 
\begin{equation}\label{metric_conf}
g_{\mu\nu} \rightarrow \Omega^2 g_{\mu\nu}\,.
\end{equation} 
In this nomenclature and related notational conventions, we have followed~\cite{Wald:1984rg}. 
We should mention that some authors call this a Weyl transformation, reserving the name `conformal transformations' for what are called `conformal isometries' in~\cite{Wald:1984rg} (for a discussion on the nomenclature, see~\cite{Karananas:2015ioa}). This transformation alters lengths of spacetime intervals, but preserves angles. The 
conformally transformed spacetime and the original one have the same causal structure. 
Since $\Omega$ is a function of spacetime, the transformation of metric affects 
different entities like the Christoffel symbols, Riemann tensor and hence the Einstein-Hilbert 
action. For gauge fields in four dimensions, the matter action remains invariant under 
conformal transformation, while for other kinds of matter fields like the scalar,
the action needs to be modified. Conformal transformation of the metric 
transforms the {Christoffel symbols} as
\begin{equation}\label{c1}
\widehat{\Gamma}^\alpha_{\mu \nu} \rightarrow \widehat{\Gamma}^\alpha_{\mu \nu}
+  \delta^\alpha_{(\mu }\widehat{\nabla}^{\phantom\beta}_{\nu )} \ln \Omega
-g_{\mu \nu} g^{\alpha \beta}\widehat{\nabla}_\beta \ln\Omega\,,
\end{equation}
where the symmetric combination is defined as $ A_{(\alpha}B_{\beta)} = A_\alpha B_\beta 
+ A_\beta B_\alpha\,$. The quantities which are defined using the torsion-free connection 
will be denoted with a hat `~$\widehat{~}\,$~'. The transformation of torsion-free Ricci 
scalar can be written as
\begin{align}\label{c2}
\nonumber
\widehat{R} &\rightarrow \Omega^{-2}\{ \widehat{R}-2(n-1)g^{\mu \nu}\widehat{\nabla}_\mu \widehat{\nabla}_\nu \ln\Omega \\
&\qquad \qquad -(n-1)(n-2)(\widehat{\nabla}_\mu \ln\Omega)(\widehat{\nabla}^\mu \ln\Omega)\}\,.
\end{align}
This is the general formula in $n$ space-time dimensions. We will be 
concerned with the case where $n=4$. 
The equation of motion of scalar field,
\begin{align}
\widehat{\nabla}_\mu \widehat{\nabla}^\mu \phi = 0\,,
\end{align}
is not covariant under conformal transformation. The remedy is to modify the equation with the addition
of a non-minimal term like the following.
\begin{equation}\label{conf-scalar-eqn}
\widehat{\nabla}_\mu \widehat{\nabla}^\mu \phi - \frac{1}{6} \widehat{R}\phi = 0\,,
\end{equation} 
where, $ \widehat{\nabla}_\mu $ is the Levi-Civita connection. Above equation can be obtained from the total action
\begin{equation}\label{csf1}
S(\phi,g)= S_{EH}[g] - \int\sqrt{-g}\,d^4x\left[\frac{1}{2}g^{\mu \nu}
\partial_\mu \phi \partial_\nu \phi + \frac{1}{12}\widehat{R}\phi^2\right]\,,
\end{equation}
where, $ S_{EH}[g] $ is the Einstein-Hilbert action. This matter part of the action is invariant under the conformal 
transformation of Eq.~(\ref{metric_conf}) provided the scalar field transforms 
as 
\begin{equation}
\phi \to \Omega^{-1}\phi\,.
\end{equation}
Variation of the action with respect to the metric produces the 
energy-momentum tensor corresponding to the conformal scalar field,
which now includes a part that depends on the geometry because of
the $\widehat{R}\phi^2$ term,
\begin{equation}\label{csf2}
\widehat{T}_{\mu\nu} = \partial_\mu\phi \partial_\nu\phi - 
\frac{1}{2} g_{\mu\nu} g^{\alpha\beta}\partial_\alpha\phi \partial_\beta \phi 
+ \frac{1}{6}\widehat{G}_{\mu\nu}\phi^2 
+ \frac{1}{6} \left[ g_{\mu\nu} \widehat{\nabla}_\sigma \widehat{\nabla}^\sigma\phi^2 
- \widehat{\nabla}_\mu \widehat{\nabla}_\nu \phi^2 \right]\,.
\end{equation}
This $ \widehat{T}_{\mu\nu} $ is a conserved tensor as expected,
\begin{equation}
\widehat{\nabla}^\mu \widehat{T}_{\mu\nu} = 0\,.
\end{equation}

In this paper we discuss the conformal transformation of the vierbein and the spin 
connection and investigate the conformal properties of the action in VEP formalism. 
In Sec.~\ref{tetrad-GR}, we go through the basics of conformal transformation conformally 
invariant scalar field and fermionic field in tetrad formulation of GR.
In Sec.~\ref{conf.VEP}, we briefly discuss the VEP formalism 
and investigate the conformal transformation of the vierbein and the spin connection. 
The spin connection is an independent variable, and thus its transformation remains
indeterminate at this stage. It is, however, possible to make different choices of 
transformations without disturbing metric compatibility. In this respect we discuss 
two such choices: one with inhomogeneously transforming torsion (Nieh-Yan theory) 
and other with invariant torsion which does not seem to have been discussed in 
literature before. In Sec.~\ref{dynamical-torsion-sec} we 
discuss the conformal properties of dynamically generated torsion with specific fields. In 
Sec.~\ref{conclusion} we summarise the results obtained in the paper. We give 
a general transformation of the 
spin connection which, in suitable limits, reduces to Nieh-Yan theory or invariant torsion.

\section{Conformal transformation in tetrad formulation}\label{tetrad-GR}

This tetrad formulation of gravity is particularly useful in describing fermionic 
fields in curved spacetime. In this formalism, the action of General Relativity is written 
in terms of vierbeins $ e^I_\mu $\,, also called tetrads, instead of the metric. The 
spacetime indices, denoted by lower-case Greek letters $\mu\,, \nu\,, \dots$, 
are raised and lowered by the spacetime metric $ g $\,, while the internal indices, denoted by 
upper-case Latin letters $A\,, B\,, \dots\,,$ are raised and lowered by the internal Minkowski metric $ \eta $. 
The internal space is a flat space associated to each point of the spacetime manifold. Tetrads can be 
thought of as linear isomorphisms between the tangent space and the internal space. Tetrads are 
related to the metric by
\begin{equation}
e^I_\mu e^J_\nu \eta_{IJ} = g_{\mu\nu}\,.
\end{equation}
We can think of this equation as describing the orthogonality of tetrads. The inverse 
tetrads, also called co-tetrads, are written as $ e^\mu_I $ and satisfy
\begin{equation}
e^\mu_I e^I_\nu=\delta^\mu_\nu\,.
\end{equation}
The  tetrad determinant  
is the same as the square root of the determinant of the metric, 
$ |e| = \sqrt{-g} $. We also define an internal connection $ D $
such that its action on any smooth section $S$ is given by
\begin{equation}
(D_\mu S)^I=\partial_\mu S^I + \omega^I_{\mu J} S^J\,,
\end{equation}
where $\omega^{I}_{\mu J}$ is a connection one form, called the spin connection. In order to identify the tetrad formulation with GR, we write the Christoffel symbols of the metric formalism using the vierbein and spin connection
\begin{equation}
\widehat{\Gamma}^\alpha_{\mu \nu}= e^\alpha_I \partial_\mu e^I_\nu +
\omega^I_{\mu J} e^J_\nu e^\alpha_I\,. 
\label{gammadef}
\end{equation}
Metric compatibility of the corresponding Levi-Civita connection enables us to 
express $\omega$ in terms of tetrads as
\begin{equation}
\omega^{IJ}_\mu =
\frac{1}{2} e_{\mu K}\left( \Theta^{KIJ} - \Theta^{IJK} - \Theta^{JKI} \right) \,,
\label{spintetrad}
\end{equation}
where for convenience we have defined the quantity $\Theta^{IJK}$ as
\begin{equation}\label{theta}
\Theta^{IJK} = e^I_\nu \left[ e^{\mu J} \partial_\mu e^{K\nu} - e^{\mu K} \partial_\mu e^{\nu J} \right].
\end{equation}
We calculate the Riemann tensor, Ricci tensor and Ricci scalar
by successive contraction with the vierbein,
\begin{align}
\widehat{R}^\rho_{~\sigma \mu \nu} &= \widehat{F}^I_{\mu \nu J} e^\rho_I e^J_\sigma\,,\\
\widehat{R}_{\sigma\nu} &= \widehat{F}^I_{\mu \nu J} e^\mu_I e^J_\sigma\,,\label{rictensor}\\
\widehat{R} &= \widehat{F}^{IJ}_{\mu \nu} e^\mu_I e^\nu_J\,.\label{ricscalar}
\end{align}
Here $ \widehat{F}^{IJ}_{\mu\nu} $ is the curvature of the connection $ D $.
The tetrad action for gravity is the Einstein-Hilbert action in which 
the Ricci scalar has been replaced by Eq.~(\ref{ricscalar}),  
and the metric determinant, by that of tetrads,
\begin{equation}\label{a4}
S_{tetrad}[e]=\frac{1}{2 \kappa}\int |e|d^4x\, \widehat{F}^{IJ}_{\mu \nu} e^\mu_I e^\nu_J\,.
\end{equation}
Here $ \kappa = 8 \pi G $. Variation of the action with respect to the tetrads produces 
the equation
\begin{equation}\label{EE.tetrad.vac}
2\widehat{F}^{IJ}_{\lambda\nu} e^\lambda_I - e^J_\nu \widehat{F}^{KL}_{\rho\sigma} e^\rho_K
e^\sigma_L = 0\,. 
\end{equation}
Contracting with $e_{\mu J}$, and using Eq.~(\ref{rictensor}), we
get the familiar form 
\begin{equation}
\widehat{R}_{\mu \nu}-\frac{1}{2}g_{\mu \nu} \widehat{R} =0\,.
\label{einstein}
\end{equation}
If we include matter fields, the tetrad action reads 
\begin{equation}
S_{Total} = \frac{1}{2\kappa} \int |e|\, d^4x\, \widehat{F}^{IJ}_{\mu\nu} e^\mu_I e^\nu_J 
+ S_M\,, 
\end{equation}
where $ S_M = \int\, |e|\, d^4x\, \mathcal{L}_M $ is the action for any matter field present. 
The equation of motion obtained by variation with respect to 
the tetrad is thus
\begin{equation}\label{EE-total-tetrad}
\widehat{F}^{IJ}_{\alpha \mu} e^\alpha_I - \frac{1}{2} e^J_\mu \widehat{F}^{KL}_{\alpha\beta} 
e^\alpha_K e^\beta_L = \kappa \widehat{T}_{\mu\alpha}e^{\alpha J}\,,
\end{equation}
where $ \widehat{T}_{\mu \alpha} $ is the usual energy-momentum tensor for the matter. 
As before, we can contract this equation with the tetrad to obtain the familiar form, 
$ \widehat{G}_{\mu\nu} = \kappa \widehat{T}_{\mu \nu} $\,. 
%
\subsection{Fermionic field in tetrad formalism}
The advantage of having the spin connection is that we can write an action for fermionic fields 
in curved spacetime. The $\gamma$-matrices are defined on the flat internal space and then brought
to the spacetime using tetrads, while the covariant derivative on the fermionic field is defined 
in terms of the spin connection. In general, the spin connection is treated as an independent variable while considering the fermionic field~\cite{Fock:1929vt,Pollock:2010zz,Hehl:1976kj,Bojowald:2007nu}. We will discuss this in Sec.~\ref{conf.VEP}. When we restrict to the torsion-free case however,
the connection is not a free variable, but $\omega^{IJ}_\mu$ of Eq.~(\ref{spintetrad}). The total 
action of gravity with a minimally coupled fermion in this case is written as~\cite{Bojowald:2007nu}
\begin{align}\label{total-action-tetrad-fermion}
S[e,A,\psi] = \frac{1}{2\kappa}S_{tetrad}[e] + \int |e| d^4x \frac{i}{2}\left[ \left( \bar{\psi}\gamma^K e^\mu_K ~^\psi \widehat{D}_\mu\psi - (\bar{\psi}\gamma^K e^\mu_K ~^\psi\widehat{D}_\mu\psi)^\dagger \right)\right]\,,
\end{align}
where $ S_{tetrad}[e] $ is the gravity action given in Eq.~(\ref{a4}). The covariant derivative $ ~^\psi\widehat{D}_\mu $ acts on the spinor $ \psi $ as
\begin{equation}
~^\psi \widehat{D}_\mu \psi = \partial_\mu \psi - \frac{i}{4} \omega^{IJ}_\mu \sigma_{IJ} \psi\,,
\end{equation}
where $\sigma_{IJ} = \frac{i}{2}\left[\gamma_I\,, \gamma_J\right]\,.$
The fermionic Lagrangian can thus be written as
\begin{equation}\label{spinor-lagrangian-tetrad-gr}
\mathcal{L}_F = \frac{i}{2}	\left(\bar{\psi}\gamma^K e^\mu_K\partial_\mu\psi - \partial_\mu\bar{\psi}\gamma^K e^\mu_K\psi - \frac{i}{4} \omega^{IJ}_\mu e^{\mu K} \bar{\psi}\{\gamma_K,\sigma_{IJ} \} \psi \right)\,.
\end{equation}
The $ \gamma $ and $ \sigma $ matrices carry internal flat space indices and have 
the usual properties for metric signature ($-+++$). We note here that for 
the choice of signature $(+---)$\,, which is popular in quantum field theory, 
we need to replace $ \gamma$  by  $i\gamma\,$ in all of these expressions.

Extremising the action of Eq.~(\ref{complete-action}) with respect to the tetrad and the fermion,
we obtain their equations of motion,
\begin{subequations}\label{complete-action-eqs-tetrad-gr}
	\begin{align}
	&\delta e^\nu_J : \widehat{R}_{\mu\nu} - \frac12 g_{\mu\nu} \widehat{R} = \kappa \widehat{T}_{\mu\nu}(\psi, \bar\psi)\,, \label{tetrad-eom}
	\\
	&\delta \bar{\psi} : 2\gamma^K e^\mu_K  \partial_\mu \psi + e^\alpha_I \partial_\mu e^I_\alpha~\gamma^K e^\mu_K \psi + \partial_\mu e^\mu_K \gamma^K \psi - \frac{i}{4}\omega^{IJ}_\mu e^{\mu K} \{\gamma_K,\sigma_{IJ} \} \psi = 0 \label{Dirac-equation-tetrad-gr}\,.
	\end{align}
\end{subequations}
In addition, by varying $\psi\,$ we get an equation which is the adjoint of Eq.~(\ref{Dirac-equation-tetrad-gr}).
Here $ \widehat{T}_{\mu\nu}(\psi, \bar\psi) $ is the symmetric and conserved energy-momentum tensor of the fermionic field, 
\begin{align}\label{fermion-em-tetrad-gr}
\widehat{T}_{\mu\nu} ({\psi,\bar{\psi}}) = \frac12 \left[	\frac{i}{2}\left( (\partial_\mu \bar{\psi}) \gamma_I \psi e^I_\nu - \bar{\psi} \gamma_I (\partial_\mu \psi) e^I_\nu + \frac{i}{4} \omega^{IJ}_\mu e^K_\nu \bar{\psi}\{\gamma_K,\sigma_{IJ}\}\psi\right) + (\mu\leftrightarrow \nu)\right]\,.
\end{align}
It should by noted that for obtaining the above expression for $ \widehat{T}_{\mu\nu} $, we
have also varied the spin connection $\omega^{IJ}_\mu$ with respect to tetrads. In fact the 
terms that come from the variation of the spin connection, along with Eq.~(\ref{Dirac-equation-tetrad-gr}), give the symmetric form of $ \widehat{T}_{\mu\nu}\, $. Eq.~(\ref{Dirac-equation-tetrad-gr}) is the Dirac equation in torsion-free curved spacetime. We can cast the Dirac equation in a familiar form by using the expression for $ \omega^{IJ}_\mu $ of Eq.~(\ref{spintetrad}),
\begin{align}
\gamma^K e^\mu_K ~^\psi \widehat{D}_\mu \psi = 0\,.
\end{align}
Let us now consider conformal transformations in the language of tetrads and spin connections.
\subsection{Conformal transformation in tetrad formulation}
We can expect that fields and their actions will transform under conformal transformations in the same way 
in the tetrad formalism as they do in the usual metric formulation of GR. 
The transformation of $ g_{\mu\nu} $ suggests that the tetrads should transform 
in the following manner,
\begin{equation}\label{tetrad_conf}
e^I_\mu \rightarrow \Omega e^I_\mu\,,
\end{equation}
while the co-tetrads should transform as
\begin{equation}\label{cotetrad-conf}
e^\mu_I \rightarrow \Omega^{-1}e^\mu_I\,.
\end{equation}
The conformal transformation of the spin connection, given by $\omega^{IJ}_\mu$ of Eq.~(\ref{spintetrad}), 
can be found from the transformation of the vierbein alone,
\begin{align}\label{ax}
\omega^I_{\mu J} &\rightarrow \omega^I_{\mu J} + e^I_\mu e^{\nu}_ J 
\partial_\nu \ln\Omega - e_{\mu J} e^{\nu I}\partial_\nu \ln \Omega\,.
\end{align}
It was argued in~\cite{Maluf:1985fj} that the above equation is 
the conformal transformation of the spin connection even in the presence of fermionic matter. 
We will, however, see  in Sec.~\ref{fermion.torsion} that in case the spin connection is
treated as an independent variable, the above transformation may not be quite correct.

Eq.~(\ref{ax}) leads to the following transformation of the Ricci scalar
\begin{align}
\widehat{F}^{IJ}_{\mu\nu}e^\mu_I e^\nu_J \rightarrow
\Omega^{-2}\left[\widehat{F}^{IJ}_{\mu\nu}e^\mu_I e^\nu_J 
- 6  \widehat{\nabla}_\mu \widehat{\nabla}^\mu \ln\Omega
- 6  \left(\widehat{\nabla}_\mu \ln\Omega\right)
\left(\widehat{\nabla}^\mu \ln\Omega\right)
\right] 
\,. 
\label{vep.conf}
\end{align}
The covariant derivative here is to be understood as being written 
in terms of $\omega\,,$ 
\begin{equation}
\widehat{\nabla}_\mu V_\nu 
= \partial_\mu V_\nu -e^\alpha_I \partial_\mu e^I_\nu V_\alpha 
- \omega^I_{\mu J} e^J_\nu e^\alpha_I V_\alpha\,.
\label{Aderiv}
\end{equation}
This is the same torsion-free covariant derivative corresponding to the Christoffel symbols written in a different form. 

Let us first discuss the conformal scalar field in tetrad formulation. We write the action of 
Eq.~(\ref{csf1}) in terms of tetrads,
\begin{align}
S[e,A,\phi] = S_{tetrad}[e,A] + \int |e|d^4x \left[-\frac12 e^\mu_I e^{\nu I} \partial_\mu \phi \partial_\nu \phi - \frac{1}{12}\widehat{F}^{IJ}_{\mu \nu} e^\mu_I e^\nu_J \phi^2\right]\,.
\label{scalar-action-tetrad}
\end{align}
We get the following set of equations by extremising the action with respect to the independent variables,
\begin{subequations}\label{conf-scalar-eqs-tetrad}
	\begin{align}
	&\delta e^\nu_J : \widehat{R}_{\mu\nu} - \frac12 g_{\mu\nu} \widehat{R} = \kappa \widehat{T}_{\mu\nu}\\
	&\delta \phi : \widehat{\nabla}_\mu \widehat{\nabla}^\mu \phi - \frac16 \widehat{R} \phi = 0\label{conf-scalar-eqs-3-tetrad}\,.
	\end{align}
\end{subequations}
Here we have contracted the equations with suitable tetrads to cast them in familiar forms. $T_{\mu\nu}$
is the same conserved energy momentum tensor obtained in Eq.~(\ref{csf2}) using the metric formulation of GR. 
The scalar field equation
above is covariant under the conformal transformation of tetrads and $\phi$. These results are expected as tetrad
formulation is nothing but GR written in a different language.

The fermionic action of Eq.~(\ref{total-action-tetrad-fermion}) is invariant under conformal transformation 
of tetrads, provided the fermionic field transforms as
\begin{align}
\left(\psi,~\bar{\psi}\right) \rightarrow \left(\Omega^{-\frac32}\psi,~\Omega^{-\frac32}\bar{\psi}\right)\,.
\label{psi-xform}
\end{align} 
In the tetrad-only formulation discussed above we have not considered torsion anywhere. Let us now go to a 
broader picture where the connection is not presumed to be torsion-free.

\section{Matter fields and conformal transformation in VEP formalism}\label{conf.VEP}
The vierbein Einstein-Palatini (VEP) formulation is a more general version of the tetrad formulation, 
involving relaxation of the torsion-free condition off-shell. 
The spin connection is considered an independent
variable in this formulation, and we will denote this by $A^I_{\mu J}$ to distinguish it 
from the torsion-free spin connection $\omega^I_{\mu J}$\,. The affine connection defined by
\begin{equation}
\Gamma^\alpha_{\mu \nu}= e^\alpha_I \partial_\mu e^I_\nu +
A^I_{\mu J} e^J_\nu e^\alpha_I\,,
\label{gammadef1}
\end{equation}
is not torsion-free i. e., 
\begin{align}
\Gamma^\alpha_{\mu \nu} \neq \Gamma^\alpha_{\nu \mu}\,.
\end{align}
In the absence of matter, the torsion-free condition appears 
from the equation of motion of the spin connection where $A^{IJ}_\mu$ is given by 
$\omega^{IJ}_\mu$ of tetrad formalism. To see this, we first 
define the VEP action. This action is the same as the Einstein-Hilbert action, but written 
in terms of tetrads and the spin connection. We define the VEP action by replacing the metric with
tetrads and the Christoffel symbols with the general affine connection of Eq.~(\ref{gammadef1}), 
\begin{equation}\label{vep-action}
S_{VEP}[e,A]=\frac{1}{2 \kappa}\int |e|d^4x\, F^{IJ}_{\mu \nu} e^\mu_I e^\nu_J\,.
\end{equation}
Extremising the action with respect to tetrad produces the same equation as before,
\begin{equation}\label{EE.tetrad.vac}
2F^{IJ}_{\lambda\nu} e^\lambda_I - e^J_\nu F^{KL}_{\rho\sigma} e^\rho_K
e^\sigma_L = 0\,. 
\end{equation}
Contracting with $e_{\mu J}$ and using Eq.~(\ref{rictensor}), we
get the familiar form 
\begin{equation}
R_{\mu \nu}-\frac{1}{2}g_{\mu \nu} R =0\,.
\label{einstein}
\end{equation}
This equation still contains the spin connection which is an unknown quantity till now. In order to find an expression for it, we vary the action with respect to the spin connection and obtain the equation
\begin{equation}\label{torsion_free}
-e_K^\alpha (\partial_\mu e^K_\alpha) e^\mu_I e^\nu_J 
- (\partial_\mu e^\mu_I) e^\nu_J - e^\mu_I (\partial_\mu e^\nu_J)
 + A^K_{\mu I} e^\mu_K e^\nu_J - A^K_{\mu I} e^\mu_J e^\nu_K = 0\,.
\end{equation}
The above equation can be solved to produce
\begin{equation}\label{torsion-free-1}
A^{IJ}_\mu = \omega^{IJ}_\mu\,,
\end{equation}
where $\omega^{IJ}_\mu$ is as was defined in Eq.~(\ref{spintetrad}). 
Then $R_{\mu\nu}$ in Eq.~(\ref{einstein}) can be replaced by $\widehat{R}_{\mu\nu}\,,$ and 
we see that even though we started with an independent connection, we have recovered 
GR on-shell. 
Eq.~(\ref{spintetrad}) is the expression for the spin connection in the absence
of matter, or more precisely in the absence of matter which couples to 
the spin connection. Thus, in vacuum the VEP formalism is equivalent on shell to the tetrad formulation.

In the presence of matter the absence of torsion is not guaranteed. Let us consider VEP action
with matter field to see this in more detail,
\begin{equation}
S_{Total} = \frac{1}{2\kappa} \int |e|\, d^4x\, F^{IJ}_{\mu\nu} e^\mu_I e^\nu_J 
+ S_M \,.
\end{equation}
Here the components of the spin connection $A^I_{\mu J}$ are again taken to be independent 
variables. 
The equation of motion, obtained by varying this action with respect to 
the tetrad, is 
\begin{equation}\label{EE-total-vep}
F^{IJ}_{\alpha \mu} e^\alpha_I - \frac{1}{2} e^J_\mu F^{KL}_{\alpha\beta} 
e^\alpha_K e^\beta_L = \kappa \Theta^J_\mu\,,
\end{equation}
where we have written 
\begin{align}
\Theta^J_\mu = -\frac{\delta S_M}{\delta e^\mu_J}\,.
\end{align}
After contraction with a suitable tetrad, we get the familiar form
\begin{align}\label{EE-tetrad-1}
	R_{\mu\nu} - \frac12 g_{\mu\nu} R = \kappa \Theta_{\mu\nu}\,.
\end{align}
We should point out that $ \Theta_{\mu \nu} = \Theta^J_\mu e_{J\nu} $ is not the usual energy-momentum tensor 
for the matter, because for $ \Theta_{\mu \nu} $ to be energy-momentum tensor, the left 
hand side of the above equation must be symmetric and torsion-free. 
If the spin connection couples to matter, the right hand side of Eq.~(\ref{torsion_free}) 
will not vanish and in general, the connection will be given as
\begin{equation}
A^{IJ}_\mu = \omega^{IJ}_\mu + \Lambda^{IJ}_\mu\,,
\label{decomp}
\end{equation}
where $ \omega^{IJ}_\mu$ is given by Eq.~(\ref{spintetrad}), and $\Lambda^{IJ}_\mu$ 
characterises the torsion part, more specifically contorsion. We note that the $ \Lambda $ term in the equation above comes from the coupling of the spin connection to the matter field considered. As a result this term will always be suppressed by $\kappa$. This will be clear once we consider specific fields. Let us now discuss how $ \Lambda $ relates to torsion. First we write the torsion tensor in terms of tetrads 
and the spin connection,
\begin{equation}\label{torsion-tetrad-spincon}
C^\alpha_{\phantom{\alpha}\mu\nu} \equiv \Gamma^\alpha_{\mu \nu} - \Gamma^\alpha_{\nu\mu} =  e^\alpha_I 
\partial_\mu e^I_\nu - e^\alpha_I \partial_\nu e^I_\mu + A^I_{\mu J} e^J_\nu e^\alpha_I - A^I_{\nu J} e^J_\mu e^\alpha_I\,.
\end{equation}
In terms of $ \Lambda $, the above expression can be written as
\begin{align}\label{torsion-def-Lambda}
C^\alpha_{\phantom{\alpha}\mu\nu} = \Lambda^I_{\mu J} e^J_\nu e^\alpha_I - \Lambda^I_{\nu J} e^J_\mu e^\alpha_I\,.
\end{align}
We know that in presence of non-zero torsion, the components of the affine connection can be written as
\begin{align}
\Gamma^\alpha_{\phantom{\alpha}\mu\nu} = \widehat{\Gamma}^\alpha_{\phantom{\alpha}\mu\nu} - S^\alpha_{\phantom{\alpha}\mu\nu}\,,
\end{align}
where $ \widehat{\Gamma}^\alpha_{\phantom{\alpha}\mu\nu} $ are the Christoffel symbols, i.e. the components of 
the Levi-Civita connection, and $ S^\alpha_{\phantom{\alpha}\mu\nu} $ is the contorsion tensor given by
\begin{align}\label{contorsion-1}
S^\alpha_{\phantom{\alpha}\mu\nu} = \frac12\left(C^{\phantom{\mu\nu}\alpha}_{\mu\nu} + C^{\phantom{\mu\nu}\alpha}_{\nu\mu}-C^\alpha_{\phantom{\alpha}\mu\nu}\right)\,.
\end{align}
Using Eq~(\ref{decomp}) and Eq.~(\ref{contorsion-1}), we see that $ \Lambda $ plays 
the role of contorsion tensor in the VEP formalism,
\begin{align}
S^\alpha_{\phantom{\alpha}\mu\nu} = \Lambda^I_{\mu J} e^J_\nu e^\alpha_I\,.
\end{align}
It is worth mentioning that it is the contorsion tensor that couples to spinors, ultimately leading 
to non-zero torsion~\cite{Hehl:1976kj}. 

Let us discuss how we can obtain the proper energy-momentum tensor on the right hand side 
of Eq.~(\ref{EE-tetrad-1}). We use the above expression in Eq.~(\ref{EE-tetrad-1}) and take all the terms coming from $\Lambda$ to the right hand side. We obtain the following equation
\begin{align}
\widehat{R}_{\mu\nu} - \frac12 g_{\mu\nu}\widehat{R} = \kappa\widehat{T}_{\mu\nu} + \mathcal{O}(\kappa^2)+...\,
\end{align}
where quantities with a hat `~$ \widehat{~} $~' over them are constructed with the torsion-free connection, as before. 
$ \widehat{T}_{\mu\nu} $ is the symmetric and conserved energy momentum tensor and higher order terms are 
contributions due to dynamically generated torsion. The conservation of the energy-momentum tensor is to be understood in terms of the torsion-free Levi-Civita connection. The procedure for obtaining a symmetric energy-momentum tensor of the spinor field, in particular, was discussed in~\cite{Hehl:1976kj}.
%

\subsection{Fermionic field in VEP formalism}
In tetrad formulation we saw how we can write fermions in gravity or more specifically torsion-free gravity. In VEP formalism fermionic field becomes more interesting because of its ability to couple to torsion. Let us consider the action to see this in detail.
\begin{align}
S[e,A,\phi,\psi] = S[e,A] + \int |e|\, d^4x\, \frac{i}{2}\left[ \left( \bar{\psi}\gamma^K e^\mu_K ~^\psi D_\mu\psi - (\bar{\psi}\gamma^K e^\mu_K ~^\psi D_\mu\psi)^\dagger \right)\right],
\label{complete-action}
\end{align}
where $ S[e,A] $ is the VEP action of Eq.~(\ref{vep-action}) and the derivative $ ~^\psi D $ acts on the spinor $\psi$ 
via the spin connection $A\,,$
\begin{equation}
~^\psi D_\mu \psi = \partial_\mu \psi - \frac{i}{4} A^{IJ}_\mu \sigma_{IJ} \psi\,.
\end{equation}
As before, the fermionic Lagrangian can be written in the following manner,
\begin{equation}\label{fermion-lagrangian-VEP}
\mathcal{L}_F = \frac{i}{2}	\left(\bar{\psi}\gamma^K e^\mu_K\partial_\mu\psi - 
\partial_\mu\bar{\psi}\gamma^K e^\mu_K\psi - \frac{i}{4} A^{IJ}_\mu e^{\mu K} 
\bar{\psi}\{\gamma_K,\sigma_{IJ} \} \psi \right)\,.
\end{equation}
Extremising the action with respect to the different variables, we get
\begin{subequations}\label{complete-action-eqs-vep}
	\begin{align}
	\delta e^\nu_J\, &: \, R_{\mu\nu} - \frac12 g_{\mu\nu} R = \frac{i\kappa}{2} \left[ \left(\partial_\nu \bar{\psi}\right) \gamma_I \psi e^I_\mu - \bar{\psi} \gamma_I (\partial_\nu \psi) e^I_\mu + \frac{i}{4} A^{IJ}_\nu e^K_\mu \bar{\psi}\{\gamma_K,\sigma_{IJ}\}\psi \right]\,,\label{EE-vep-1}\\
	\delta A^{IJ}_\nu\, &: \, A^{IJ}_\mu = \omega^{IJ}_\mu [e] + \frac{\kappa}{8}\bar{\psi}\{\gamma_K,\sigma^{IJ} \}\psi e^K_\mu\,, \label{complete-action-eqs-2}\\
	\delta \bar{\psi}\, &: \, 2\gamma^K e^\mu_K  \partial_\mu \psi + e^\alpha_I \partial_\mu e^I_\alpha~\gamma^K e^\mu_K \psi + \partial_\mu e^\mu_K \gamma^K \psi - \frac{i}{4}A^{IJ}_\mu e^{\mu K} \{\gamma_K,\sigma_{IJ} \} \psi = 0 \label{Dirac-equation-vep-1}\,.
	\end{align}
\end{subequations}
We can simplify Eq.~(\ref{Dirac-equation-vep-1}) by using identities of $ \gamma $ and $ \sigma $ matrices; and the definition of torsion in Eq.~(\ref{torsion-tetrad-spincon}),
\begin{align}\label{Dirac-eq-wo-skew}
\gamma^K e^\mu_K  \partial_\mu \psi + \frac12 C^\alpha_{\phantom{\alpha}\mu\alpha} e^\mu_K \gamma^K \psi - \frac{i}{4}A^{IJ}_\mu e^{\mu K} \gamma_K\sigma_{IJ} \psi = 0.
\end{align}
Let us now see what happens to the above equation if we use the on-shell expression of the spin connection given by Eq.~(\ref{complete-action-eqs-2}). The first term in expression is exactly same as in Eq.~(\ref{spintetrad}). The last term has appeared due to the fermionic field. This term  can be identified with the contorsion $ \Lambda^{IJ}_\mu $ as given in Eq.~(\ref{decomp}),
\begin{align}\label{lambda-vep}
\Lambda^{IJ}_\mu = \frac{\kappa}{8}\bar{\psi}\{\gamma_K,\sigma^{IJ} \}\psi e^K_\mu\,.
\end{align}
We can use the following identity
\begin{align}
	\{\sigma^{IJ},\gamma^K\} = 2 \epsilon^{IJKL} \gamma_L \gamma_5\,,
\end{align}
to write $\Lambda^{IJ}_\mu$ as
\begin{align}
\Lambda^{IJ}_\mu = \frac{\kappa}{4} \epsilon^{IJKL}\bar{\psi}\gamma_L\gamma_5\psi e_{K\mu}\,.
\end{align}
This term results in the following non-vanishing expression for the on-shell torsion tensor:
\begin{align}\label{torsion}
~^{OS}C^\alpha_{\phantom{\alpha}\mu\nu} = \frac{\kappa}{2} \epsilon^{IJKL} \bar{\psi} \gamma_L \gamma_5 \psi e^\alpha_I e_{J\mu} e_{K\nu}\,.
\end{align}
Clearly, on-shell torsion, generated by a fermion source, is totally antisymmetric. 
If we identify $ C^\alpha_{\phantom{\alpha}\mu\alpha} $ of Eq.~(\ref{Dirac-eq-wo-skew}) with $ ~^{OS}C^\alpha_{\phantom{\alpha}\mu\nu} $ we can see that the second term in the equation goes away due to the total antisymmetry of the torsion tensor given by Eq.~(\ref{torsion}). Thus we are left with the following equation.
\begin{align}\label{Dirac-eq-simp}
\gamma^K e^\mu_K  \partial_\mu \psi - \frac{i}{4}A^{IJ}_\mu e^{\mu K} \gamma_K \sigma_{IJ} \psi = 0\,
\end{align}
or simply
\begin{align}\label{spinor-eq-vep-on-shell}
\gamma^K e^\mu_K ~^\psi D_\mu \psi = 0\,.
\end{align}
This looks like the equation we obtained for fermionic field in tetrad formulation, but it is different in the sense that the spin connection, and thus the derivative, now contain non-zero torsion. If we now use Eq.~(\ref{complete-action-eqs-2}) for the on-shell expression of the spin connection, we get a non-linear spinor equation with cubic term resulting from torsion, as has been noted in~\cite{Hehl:1976kj},
\begin{align}\label{NLD}
\gamma^K e^\mu_K  \partial_\mu \psi - \frac{i}{4}\omega^{IJ}_\mu e^{\mu K} \gamma_K \sigma_{IJ} \psi -\frac{i\kappa}{64}\bar{\psi}\{\gamma_K,\sigma_{IJ}\}\psi\{\gamma^K,\sigma^{IJ}\}\psi  = 0\,.
\end{align}
This extra term is suppressed by a factor of $\kappa\,.$ 

Let us also use the on-shell expression of the spin connection given by Eq.~(\ref{complete-action-eqs-2}), and the Dirac equation of Eq.~(\ref{Dirac-equation-vep-1}) in Eq.~(\ref{EE-vep-1}). After some lengthy but straightforward calculations 
we get the following equation. 
\begin{align}\label{EE-spinor-sym}
\widehat{R}_{\mu\nu} - \frac12 g_{\mu\nu} \widehat{R} = \kappa \widehat{T}_{\mu\nu} ({\psi,\bar{\psi}}) -\frac{3 \kappa^2}{16} g_{\mu\nu} \bar{\psi}\gamma_I\gamma_5 \psi \bar{\psi} \gamma^I\gamma_5\psi  \,.
\end{align}
Here, $ \widehat{T}_{\mu\nu} ({\psi,\bar{\psi}}) $ is the symmetric and conserved energy-momentum that comes from the torsion-free matter Lagrangian as obtained in Eq.~(\ref{fermion-em-tetrad-gr}), which comes from a torsion-free theory of gravity. The conservation of $ T_{\mu\nu} ({\psi,\bar{\psi}}) $ is to be understood in terms of the torsion-free derivative operator $ \widehat{\nabla} $. The additional term of $ \mathcal{O}(\kappa^2) $ has appeared due to torsion. Using generalised Fierz identities for the spinor field~\cite{Takahashi:1982bb, Pal:2007dc, Nieves:2003in}, we can write Einstein equations as
\begin{align}
\widehat{R}_{\mu\nu} - \frac12 g_{\mu\nu} \widehat{R} = \kappa \widehat{T}_{\mu\nu} ({\psi,\bar{\psi}}) - g_{\mu\nu}\frac{3 \kappa^2}{16}\left( (\bar{\psi}\gamma_5\psi)^2 -(\bar{\psi} \psi)^2  \right)\,.
\end{align}
Let us conclude this section with a discussion on the non-linear Dirac Eq.~(\ref{NLD}). As already mentioned, the effect of torsion on the equation is realised through the cubic term. We can write 
an effective Lagrangian without torsion that gives us the same result as obtained above. 
To do this we need to modify the torsion-free spinor Lagrangian of~(\ref{spinor-lagrangian-tetrad-gr}) with the addition of a quartic term as
\begin{align}
\mathcal{L}_F = \frac{i}{2}	\left(\bar{\psi}\gamma^K e^\mu_K\partial_\mu\psi - \partial_\mu\bar{\psi}\gamma^K e^\mu_K\psi - \frac{i}{4} \omega^{IJ}_\mu e^{\mu K} \bar{\psi}\{\gamma_K,\sigma_{IJ} \} \psi -\frac{i\kappa}{64}\bar{\psi}\{\gamma_K,\sigma_{IJ}\}\psi\bar{\psi}\{\gamma^K,\sigma^{IJ}\}\psi \right)\,.
\end{align}
We can write the above Lagrangian in a better form using Fierz identities as
\begin{align}
\mathcal{L}_F = \frac{i}{2}	\left(\bar{\psi}\gamma^K e^\mu_K\partial_\mu\psi - \partial_\mu\bar{\psi}\gamma^K e^\mu_K\psi - \frac{i}{2}\epsilon_{IJKL} \omega^{IJ}_\mu e^{\mu K} \bar{\psi}\gamma^L\gamma_5 \psi -\frac{3 i\kappa}{8}\left((\bar{\psi}\gamma_5\psi)^2 -(\bar{\psi} \psi)^2\right) \right)\,.
\end{align}
This Lagrangian also produces the $\mathcal{O}(\kappa^2)$ term appearing in the Einstein equation.

\subsection{Conformal transformation in VEP formalism}
Let us now investigate conformal transformations in the VEP formalism. Since the connection is treated an 
independent variable except when it is on-shell, we need to discuss the conformal properties in two different 
levels: off-shell and on-shell. By on-shell, we mean that the spin connection and torsion have been replaced 
by their expressions obtained from the equations of motion.
We will discuss this in Sec.~\ref{dynamical-torsion-sec}. In the level of action (off-shell), however, we 
cannot use the on-shell equations. Nevertheless, tetrads and co-tetrads are related to the metric and they 
transform in the same way as given in Eq.~(\ref{tetrad_conf}) and Eq.~(\ref{cotetrad-conf}). 
\begin{equation}
e^I_\mu \rightarrow \Omega e^I_\mu,\qquad e^\mu_I \rightarrow \Omega^{-1}e^\mu_I\,.
\end{equation}
For the spin connection $A^{IJ}_\mu\,,$ we do not know a priori how it transforms. To find its 
transformation properties, let us write the transformation of the connection coefficients
$ \Gamma^\alpha_{\mu \nu} $ 
in the VEP formalism, which can be obtained by defining $\tilde{\Gamma}$ in
terms of the transformed tetrads and spin connection,
\begin{align}
\tilde{\Gamma}^\alpha_{\mu \nu} &= \tilde{e}^\alpha_I 
\partial_\mu \tilde{e}^I_\nu + \tilde{A}^I_{\mu J} 
\tilde{e}^J_\nu \tilde{e}^\alpha_I \nonumber\\
&= \delta^\alpha_\nu \partial_\mu (\ln\Omega) + 
e^\alpha_I \partial_\mu e^I_\nu + \tilde{A}^I_{\mu J} 
e^J_\nu e^\alpha_I\,.
\label{gamma.conf}
\end{align}
We can try to determine the transformation of ${A}^I_{\mu J} $  
from this, by positing that the corresponding connection $ \tilde{\nabla} $
is compatible with the transformed metric $\tilde{g}_{\mu \nu}\,.$
Using Eq.~(\ref{gamma.conf}) we can write
\begin{align}
\nonumber
\tilde{\nabla}_\mu \tilde{g}_{\alpha\beta} &= 
\partial_\mu \tilde{g}_{\alpha\beta} - 
\tilde{\Gamma}^\nu_{\mu \alpha} \tilde{g}_{\nu \beta} - 
\tilde{\Gamma}^\nu_{\mu \beta} \tilde{g}_{\alpha \nu} \\
\notag
&= \partial_\mu (\Omega^2 g_{\alpha\beta}) - 2 \Omega^2 \partial_\mu(\ln\Omega) g_{\alpha\beta} 
-\Omega^2( \partial_\mu e_{I(\alpha}) e^I_{\beta)} - \Omega^2\tilde{A}^{IJ}_\mu (e_{J (\alpha } e_{\beta) I}) \notag\\
&= 0\,.
\end{align}
In the last equality, we have used the orthonormality of the tetrads 
and the antisymmetry of the spin connection.
Quite clearly, antisymmetry of $ \tilde{A}^{IJ}_\mu $ in $IJ$ 
is sufficient to guarantee metric compatibility. 
Therefore metric compatibility is not sufficient to 
determine the transformation of $ A\,, $ and only shows antisymmetry 
in the internal indices $ I\,,J \,.$ We are thus in liberty to choose the 
transformation of the spin connection as long as metricity is satisfied. 
However, the different possible choices are not guaranteed to reproduce usual GR even in the 
absence of torsion. Let us now demonstrate a couple of such choices. We will also 
consider matter fields to demonstrate how these choices affect their conformal properties.

\subsection{Nieh-Yan theory}
Nieh-Yan theory~\cite{Nieh:1981xk} involves one of the possible choices of conformal transformations of the 
spin connection. We first note that although the spin connection is an independent variable, we can always 
decompose it in terms of the torsion-free $\omega\,,$ which is completely determined by the tetrads, 
and the contorsion tensor $\Lambda$ as shown in Eq.~(\ref{decomp}),
\begin{align}\label{A-decomp}
A^{IJ}_\mu = \omega^{IJ}_\mu + \Lambda^{IJ}_\mu\,.
\end{align}
In order to find the conformal transformation of $A^{IJ}_\mu$, we note that $\omega^{IJ}_\mu$ is defined 
completely in terms of the tetrads regardless of whether we consider it on-shell or of-shell, and its
transformation is given by Eq.~(\ref{ax}). The independent quantity in the connection is the contorsion 
component $\Lambda^{IJ}_\mu$ for which we do not know how it transforms off-shell. 
In other words, we do not have any information about torsion and its transformation. 
But as discussed above, we can make different choices about the conformal transformation of spin connection 
as long as metric compatibility is retained. The choice will dictate what physical results we get 
and also specify the transformation of torsion. The simplest choice in this regard was considered by the authors 
in~\cite{Nieh:1981xk}. They considered the spin connection to be invariant under conformal transformation,
\begin{align}\label{spin-con-trans-nieh}
	A^{IJ}_\mu \rightarrow A^{IJ}_\mu.
\end{align}
Invariance of the spin connection implies that unlike in GR, here the Riemann tensor and the Ricci tensor 
and remain invariant, while the Ricci scalar transforms homogeneously,
\begin{align}\label{Riemann-conf-nieh}
	 R^\rho_{~\sigma \mu \nu} &= F^I_{\mu \nu J} e^\rho_I e^J_\sigma \rightarrow F^I_{\mu \nu J} e^\rho_I e^J_\sigma\,,\\
	 R_{\mu\nu} &= F^I_{\sigma \mu J} e^\sigma_I e^J_\nu \rightarrow F^I_{\sigma \mu J} e^\sigma_I e^J_\nu\,,\\
	 R &= F^{IJ}_{\mu\nu} e^\mu_I e^\nu_J \rightarrow \Omega^{-2} F^{IJ}_{\mu\nu} e^\mu_I e^\nu_J \,.
	\label{Ricci-NY-conf}
\end{align}
This is one of the advantages of Nieh-Yan theory, we get a conformally covariant theory of gravity.
Let us investigate further what the invariance of spin connection implies and how it fixes the 
transformation of the torsion tensor. We note that in order for the spin connection $ A^{IJ}_\mu $ of Eq.~(\ref{A-decomp}) to remain invariant, the transformation of contorsion part $ \Lambda $ must cancel that of the torsion-free part $ \omega $ given by Eq.~(\ref{ax}), i.e.,
\begin{align}
\Lambda^{IJ}_\mu \rightarrow \Lambda^{IJ}_\mu - (e^I_\mu e^{\nu}_ J - e_{\mu J} e^{\nu I})\partial_\nu \ln \Omega\,.
\end{align} 
The transformation of torsion tensor, which is given in terms of $\Lambda$ in Eq.~(\ref{torsion-def-Lambda}), 
is thus 
\begin{align}\label{torsion-conf-nieh}
^{NY}C^\alpha_{\phantom{\alpha}\mu\nu} \rightarrow ^{NY}C^\alpha_{\phantom{\alpha}\mu\nu} 
+ \delta^\alpha_\nu \partial_\mu\ln \Omega - \delta^\alpha_\mu \partial_\nu\ln \Omega \,.
\end{align}
Clearly the Nieh-Yan torsion tensor $ ^{NY}C^\alpha_{\phantom{\alpha}\mu\nu} $ transforms inhomogeneously, or in other words, torsion acts as a gauge transformation in the conformal transformation group, which is a fundamental result of this theory. Let us now consider the fermionic field and the scalar field in Nieh-Yan theory.
%
\subsubsection{Fermionic field in Nieh-Yan theory}
Because we are working with actions, we consider the fermionic Lagrangian rather than the Dirac equation. 
Let us recall the Lagrangian of the fermionic field given in Eq.~(\ref{fermion-lagrangian-VEP}),
\begin{align}
	\mathcal{L}_F = \frac{i}{2}	\left(\bar{\psi}\gamma^K e^\mu_K\partial_\mu\psi - 
	\partial_\mu\bar{\psi}\gamma^K e^\mu_K\psi - \frac{i}{4} A^{IJ}_\mu e^{\mu K} \bar{\psi}\{\gamma_K,\sigma_{IJ} \} \psi \right)\,.
\end{align}
It can be readily seen that the Lagrangian transforms homogeneously with conformal weight of $-4$ if the fermion $\psi$ is taken to  transform in the same way as in Eq.~(\ref{psi-xform}), provided $A^{IJ}_\mu$ is invariant as in Nieh-Yan theory.
We thus expect the Dirac equation to remain conformally covariant too. But there are certain problems with the equation as we will see now. Let us recall the Dirac equation in the VEP formalism, which is given in Eq.~(\ref{spinor-eq-vep-on-shell}),
\begin{align}
\gamma^K e^\mu_K ~^\psi D_\mu \psi = 0\,.
\end{align}
If the spin connection remains invariant under conformal transformations, this equation transforms to 
\begin{align}
\Omega^{-\frac52}\left( \gamma^K e^\mu_K ~^\psi D_\mu \psi -\frac12 \gamma^K e^\mu_K \psi \partial_\mu \ln \Omega \right) = 0\,.
\end{align}
Clearly, the equation is not covariant under these transformations. The source of this problem lies in the fact that the spinor 
equation as written here was obtained after using the total antisymmetry of the on-shell torsion. 
Let us see how this equation transforms if we do not use any on-shell property of torsion that is derived from the equations of motion.
In this case we need to consider Eq.~(\ref{Dirac-eq-wo-skew}), i.e.,
\begin{align}
\gamma^K e^\mu_K  \partial_\mu \psi + \frac12 C^\alpha_{\phantom{\alpha}\mu\alpha} e^\mu_K \gamma^K \psi - \frac{i}{4}A^{IJ}_\mu e^{\mu K} \gamma_K\sigma_{IJ} \psi = 0\,.
\end{align}
If we replace $ C^\alpha_{\phantom{\alpha}\mu\alpha} $ with Nieh-Yan torsion $ ^{NY}C^\alpha_{\phantom{\alpha}\mu\alpha} $, the above equation becomes
\begin{align}
\gamma^K e^\mu_K  \partial_\mu \psi + \frac12~^{NY}C^\alpha_{\phantom{\alpha}\mu\alpha} e^\mu_K \gamma^K \psi - \frac{i}{4}A^{IJ}_\mu e^{\mu K} \gamma_K\sigma_{IJ} \psi = 0\,.
\end{align}

Now, the transformation of $^{NY}C^\alpha_{\phantom{\alpha}\mu\alpha}$ can be obtained from Eq.~(\ref{torsion-conf-nieh}) by tracing over first and the last index of the torsion tensor.
\begin{align}\label{tor-nieh-con-trace}
^{NY}C^\alpha_{\phantom{\alpha}\mu\alpha} \rightarrow ^{NY}C^\alpha_{\phantom{\alpha}\mu\alpha} + 3\partial_\mu \ln \Omega\,.
\end{align}
If we apply conformal transformations after taking the above into account, we find that the Dirac equation remains conformally covariant. It is thus clear that if we use the on-shell properties of torsion, the Dirac equations does not remain invariant under the assumptions of Nieh-Yan theory. We will see in Sec.~\ref{dynamical-torsion-sec} that on-shell torsion does not have the transformation as given in Eq.\ref{tor-nieh-con-trace} above.
%

\subsubsection{Conformal scalar in Nieh-Yan theory}
Let us start by writing the Lagrangian of the conformal scalar field in terms of VEP variables,
\begin{align}
\mathcal{L}_{\phi} = -\frac12 e^\mu_I e^{\nu I} \partial_\mu \phi \partial_\nu \phi - \frac{1}{12}F^{IJ}_{\mu \nu} e^\mu_I e^\nu_J \phi^2\,.
\end{align}
In Nieh-Yan theory it is the torsion that transforms inhomogeneouly as shown in Eq.~(\ref{torsion-conf-nieh}), and not the Ricci scalar which transforms as in Eq.~(\ref{Ricci-NY-conf}). This above Lagrangian is not covariant as a result,
with $\phi\rightarrow \Omega^{-1}\phi $\,.
The Lagrangian can be rewritten with the torsion-free part of the Ricci scalar in order to make it conformally covariant,
\begin{align}
\notag
\mathcal{L}_{\phi} = -\frac12 e^\mu_I e^{\nu I} \partial_\mu \phi \partial_\nu \phi - \frac{1}{12}F^{IJ}_{\mu \nu} e^\mu_I e^\nu_J \phi^2 - \phi^2\left(\frac16 \widehat{\nabla}_\mu ~^{NY}C^{\alpha\phantom{\alpha}\mu}_{\phantom{\alpha}\alpha} + \frac{1}{12} ~^{NY}C^\mu_{\phantom{\mu}\mu\sigma}
~^{NY}C^{\nu\phantom{\nu}\sigma}_{\phantom{\nu}\nu}\right.\\
\left.-\frac{1}{48} ~^{NY}C^{\mu\nu\sigma}~^{NY}C_{\mu\nu\sigma} 
- \frac{1}{24} ~^{NY}C^{\mu\nu\sigma}~^{NY}C_{\nu\mu\sigma}\right)\,.
\end{align}
We can see that the coefficients of $ \phi^2 $ in the above add up to produce the Ricci scalar $\widehat{R}$ corresponding to the torsion-free connection. We can thus write the Lagrangian as
\begin{align}
\mathcal{L}_{\phi} = -\frac12 e^\mu_I e^{\nu I} \partial_\mu \phi \partial_\nu \phi - \frac{1}{12}\widehat{F}^{IJ}_{\mu \nu} e^\mu_I e^\nu_J \phi^2
\end{align}
This is the same Lagrangian of conformal scalar of Eq.~(\ref{scalar-action-tetrad}) in tetrad formulation and transforms homogeneously with conformal weight of $-4$. Evidently the corresponding equation remains conformally invariant.

\subsection{Conformally invariant torsion}
Although Nieh-Yan theory gives a conformally covariant theory of gravity, not all its results can be identified with those in Einstein's GR in the absence of torsion. We are interested in a formalism which resembles Einstein gravity and in which different entities transform in the same way as in usual GR when if torsion is taken to vanish. In such a theory, we must assume that torsion transforms homogeneously under conformal transformation, unlike in~\cite{Nieh:1981xk} where torsion acts a gauge transformation in conformal group. Here we take the spin connection $ A^{IJ}_\mu $ to transform
in the same way as $ \omega^{IJ}_\mu $ while $ \Lambda^{IJ}_\mu $ remains invariant, i.e.,
\begin{align}\label{spin-con-inv-torsion}
	A^{IJ}_\mu (\equiv\omega^{IJ}_\mu + \Lambda^{IJ}_\mu) \rightarrow A^{IJ}_\mu + (e^I_\mu e^{J\alpha} - e^J_\mu e^{I\alpha})\partial_\alpha\ln\Omega.
\end{align}
and
\begin{align}\label{lambda-inv-torsion}
\Lambda^{IJ}_\mu \rightarrow \Lambda^{IJ}_\mu\,.
\end{align}
In this case we are in torsion spacetime but conformal transformation is given by that of tetrads only. Above transformations immediately imply that torsion tensor given by the simplified form in Eq.~(\ref{torsion-def-Lambda}), remains invariant i.e.,
\begin{align}\label{Inv-torsion-conf}
	~^{Inv}C^\alpha_{\phantom{\alpha}\mu \nu} \rightarrow ~^{Inv}C^\alpha_{\phantom{\alpha}\mu \nu}\,.
\end{align}

It should be noted that if any index of the torsion tensor is raised or lowered, there 
will be a conformal weight factor due to the metric involved in the raising or lowering. 
The Ricci scalar contains extra terms involving torsion when transformed,
\begin{align}\label{Ricci-scalar-gen-trans}
	F^{IJ}_{\mu\nu}e^\mu_I e^\nu_J\rightarrow
	\Omega^{-2}\left[F^{IJ}_{\mu\nu}e^\mu_I e^\nu_J 
	- 6  \widehat{\nabla}_\mu \widehat{\nabla}^\mu \ln\Omega
	- 6  \left(\widehat{\nabla}_\mu \ln\Omega\right)
	\left(\widehat{\nabla}^\mu \ln\Omega \right) - 2 C^\alpha_{\phantom{\alpha}\alpha\mu}\widehat{\nabla}^\mu\ln \Omega
	\right]\,. 
\end{align}
It is interesting to note that although torsion itself does not transform, it couples to the transformation. We now look at the fermionic and scalar field in the presence of invariant torsion.

\subsubsection{Conformal properties of fermionic field with invariant torsion}
Let us recall the  Lagrangian of the fermionic field,
\begin{align}
\mathcal{L}_F = \frac{i}{2}	\left(\bar{\psi}\gamma^K e^\mu_K\partial_\mu\psi - \partial_\mu\bar{\psi}\gamma^K e^\mu_K\psi - \frac{i}{4} A^{IJ}_\mu e^{\mu K} \bar{\psi}\{\gamma_K,\sigma_{IJ} \} \psi \right).
\end{align}
In order to see how this Lagrangian transforms, we note that, with the transformation of the spin connection corresponding to invariant torsion, the last term in the above Lagrangian remains unchanged.
The transformations of the other two terms cancel each other and as a result, we find that the Lagrangian transforms homogeneously with conformal weight of $-4$ as before. Let us now see what the invariant torsion implies the covariance of the Dirac equation,
\begin{align}
\gamma^K e^\mu_K ~^\psi D_\mu \psi = 0\,.
\end{align}
With the spin connection transforming inhomogeneously as in Eq.~(\ref{spin-con-inv-torsion}), we can see that the corresponding term in the Dirac equation transforms as
\begin{align}
A^{IJ}_\mu e^{\mu K} \gamma_K \sigma_{IJ} \psi \rightarrow \Omega^{-\frac{5}{2}}(A^{IJ}_\mu e^{\mu K} \gamma_K \sigma_{IJ} \psi + 2i e^{\mu K} \gamma_K \psi \partial_\mu \ln \Omega)\,.
\label{Dirac-eq-A-xform}
\end{align}
This implies that the Dirac equation remains covariant, although the equation we have considered here uses the skew symmetry of on-shell torsion. This is a difference with what was observed in Nieh-Yan theory, where we showed that we could not use any on-shell property of torsion.
We also mention that the above equation transforms in the same way as the Dirac equation
in the absence of torsion as given in  Eq.~(\ref{Dirac-equation-tetrad-gr}).
In other words, the equation remains conformally covariant with 
$\psi\rightarrow\Omega^{-\frac32}\psi$ as before.

\subsubsection{Conformal scalar with invariant torsion}
Let us now see the conformal scalar with invariant torsion. As in Nieh-Yan theory we start by writing the Lagrangian of the conformal scalar in terms of the VEP variables,
\begin{align}
\mathcal{L}_{\phi} = -\frac12 e^\mu_I e^{\nu I} \partial_\mu \phi \partial_\nu \phi - \frac{1}{12}F^{IJ}_{\mu \nu} e^\mu_I e^\nu_J \phi^2\,.
\end{align}
Using the transformation of the Ricci scalar given by Eq.~(\ref{Ricci-scalar-gen-trans}), we see that the above Lagrangian transforms as
\begin{align}
\mathcal{L}_{\phi} \quad\rightarrow\quad \Omega^{-4}\mathcal{L}_{\phi} + \frac16\Omega^{-4}~^{Inv}C^\alpha_{\phantom{\alpha}\alpha\mu} \phi^2\nabla^\mu\ln \Omega\,.
\end{align}
The minimal modification that makes the Lagrangian conformally covariant is the addition of a torsion term,
\begin{align}
\mathcal{L}_{\phi} = -\frac12 e^\mu_I e^{\nu I} \partial_\mu \phi \partial_\nu \phi - \frac{1}{12}F^{IJ}_{\mu \nu} e^\mu_I e^\nu_J \phi^2 - \frac16 \phi^2 \widehat{\nabla}_\mu ~^{Inv}C^{\alpha\phantom{\alpha}\mu}_{\phantom{\alpha}\alpha}\,.
\end{align}
This Lagrangian transforms homogeneously because the added term has the following transformation.
\begin{align}
\frac16 \phi^2 \widehat{\nabla}_\mu ~^{Inv}C^{\alpha\phantom{\alpha}\mu}_{\phantom{\alpha}\alpha}
\quad \rightarrow \quad
\Omega^{-4}\left(\frac13 \phi^2 \widehat{\nabla}_\mu ~^{Inv}C^{\alpha\phantom{\alpha}\mu}_{\phantom{\alpha}\alpha}
+ \frac16 \phi^2 ~^{Inv}C^\alpha_{\phantom{\alpha}\alpha\mu}\widehat{\nabla}^\mu \ln\Omega \right)\,.
\end{align}
The corresponding scalar equation
\begin{align}
\widehat{\nabla}_\mu \widehat{\nabla}^\mu - \frac16 F^{IJ}_{\mu \nu} e^\mu_I e^\nu_J \phi - \frac13 \phi \widehat{\nabla}_\mu ~^{Inv}C^{\alpha\phantom{\alpha}\mu}_{\phantom{\alpha}\alpha}= 0\,.
\end{align}
is invariant under conformal transformation. We note that, we can also write the torsion-free part of Ricci scalar with quadratic torsion terms, as we did in case of Nieh-Yan theory, without affecting the covariance of the Lagrangian or the equation.
We can thus say that the general Lagrangian of conformally covariant scalar field is
\begin{align}
\mathcal{L}_{\phi} = -\frac12 e^\mu_I e^{\nu I} \partial_\mu \phi \partial_\nu \phi &- \frac{1}{12}F^{IJ}_{\mu \nu} e^\mu_I e^\nu_J \phi^2 
\notag\\ &
- \phi^2\left(\frac16 \widehat{\nabla}_\mu C^{\alpha\phantom{\alpha}\mu}_{\phantom{\alpha}\alpha}
+ \frac{1}{12} C^\mu_{\phantom{\mu}\mu\sigma}
C^{\nu\phantom{\nu}\sigma}_{\phantom{\nu}\nu} - 
\frac{1}{48} C^{\mu\nu\sigma}C_{\mu\nu\sigma} 
- \frac{1}{24} C^{\mu\nu\sigma}C_{\nu\mu\sigma}\right)\,.
\end{align}
Similar to what was obtained in Nieh-Yan theory, we can identify the coefficients of $\phi^2$ as $-\frac{1}{12}\widehat{R}\,$. We are thus dealing with the conformally covariant Lagrangian that we had in the tetrad formulation of GR and the corresponding scalar equation remains invariant under conformal transformations as a result.

\section{Dynamically generated torsion and conformal transformation}\label{dynamical-torsion-sec}
In this section we deal with dynamically generated (on-shell) torsion and see its effects on conformal properties of matter fields. Torsion is considered to be sourced from other dynamical fields and it comes from the equations of the spin connection. 
It is known that spinor field can produce torsion. We find that non-minimal
scalar fields can also produce torsion on-shell. But there are problems with the
conformal weight of the torsion terms if they are considered to be purely dynamically generated. We will see that on-shell torsion transforms homogeneously but unlike invariant torsion, its transforms with an overall weight. Let us demonstrate
these problems and possible solutions with specific fields.
%

\subsection{Dynamically generated torsion and conformal scalar}\label{scalar}
We write the total action, including that of the conformal scalar, in terms of VEP variables
\begin{align}
S[e,A,\phi] = S_{VEP}[e,A] + \int |e|d^4x \left[-\frac12 e^\mu_I e^{\nu I} \partial_\mu \phi \partial_\nu \phi - \frac{1}{12}F^{IJ}_{\mu \nu} e^\mu_I e^\nu_J \phi^2\right]\,.
\label{scalar-action-VEP}
\end{align}
We get three sets of equation by extremising the action with respect to the independent variables.
\begin{subequations}\label{conf-scalar-eqs}
\begin{align}
	\delta e^\nu_J\, &: \,F^{IJ}_{\alpha \mu} e^\alpha_I e_{\nu J} - \frac12 g_{\mu\nu} F^{IJ}_{\alpha \beta} e^\alpha_I e^\beta_J = \kappa \left( \partial_\mu \phi \partial_\nu \phi - \frac12 g_{\mu\nu} g^{\alpha\beta} \partial_\alpha \phi \partial_\beta \phi + \frac16  \left(F^{IJ}_{\alpha \mu} e^\alpha_I e_{\nu J} - \frac12 g_{\mu\nu} F^{IJ}_{\alpha \beta} e^\alpha_I e^\beta_J\right) \right),\label{conf-scalar-eqs-1}\\
	\delta A^{IJ}_\nu\, &: \,A^{IJ}_\mu = \omega^{IJ}_\mu [e] + \frac12 \left( e^I_\mu e^{J\alpha} - e^J_\mu e^{I\alpha} \right) \partial_\mu \ln \left( 1-\frac{\kappa \phi^2}{6} \right),\label{conf-scalar-eqs-2}\\
	\delta \phi\, &: \,\widehat{\nabla}_\mu \widehat{\nabla}^\mu \phi - \frac16 F^{IJ}_{\mu \nu} e^\mu_I e^\nu_J \phi = 0\label{conf-scalar-eqs-3}\,.
\end{align}
\end{subequations}
Here we have contracted the equations with tetrads in order to convert them to familiar forms. 
It should be noted that in Eq.~(\ref{conf-scalar-eqs-3}) the covariant derivatives in the first 
term can be chosen to be torsion-free, since
\begin{align}
\widehat{\nabla}_\mu \widehat{\nabla}^\mu \phi - \nabla_\mu \nabla^\mu \phi = S^{\alpha\mu}_{\phantom{\alpha}\phantom{\mu}\alpha} \partial_\mu \phi = 0\,,
\end{align}
which follows from the definition of contorsion tensor given in Eq.~(\ref{contorsion-1}). 
The right hand side of Eq.~(\ref{conf-scalar-eqs-1}) does not contain the full energy-momentum tensor as can be seen on comparison with Eq.~(\ref{csf2}). As a result it will not be a conserved tensor. This is because in this case the Einstein tensor $ G_{\mu\nu}\,, $ 
which appears on the left hand side of the equation,
is torsionful. We can separate the torsion-free and torsionful parts by considering the on-shell
expression of the spin connection (\ref{conf-scalar-eqs-2}). Before getting in to this, let us 
first see how Eq.~(\ref{conf-scalar-eqs-2}) can lead to non-zero torsion. The spin connection, 
up to $ \mathcal{O}(\kappa) $ can be written as
\begin{align}\label{spin-con-scalar-approx}
	A^{IJ}_\mu \approx \omega^{IJ}_\mu [e] - \frac{\kappa}{12} \left( e^I_\mu e^{J\alpha} - e^J_\mu e^{I\alpha} \right) \partial_\mu  \phi^2\,.
\end{align}

Comparing with Eq.~(\ref{decomp}) we can identify the contorsion $ \Lambda $  with the second term in this equation, 
\begin{align}\label{lambda-conf-scalar}
\Lambda^{IJ}_\mu = - \frac{\kappa}{12} \left( e^I_\mu e^{J\alpha} - e^J_\mu e^{I\alpha} \right) \partial_\mu  \phi^2\,.
\end{align}
This gives the following expression for the on-shell torsion tensor 
\begin{align}
	~^{OS}C^\alpha_{\phantom{\alpha}\mu\nu} = \frac{\kappa}{12}\left( \delta^\alpha_\mu \partial_\nu \phi^2 - \delta^\alpha_\nu \partial_\mu \phi^2 \right)\,.
	\label{scalar-OS-torsion}
\end{align}
We note here that the expressions of on-shell torsion found in Eq.~(\ref{torsion}) and Eq.~(\ref{scalar-OS-torsion}) seem to be generic, and have been found in other contexts~\cite{Fabbri:2011jp}.
Inserting Eq.~(\ref{spin-con-scalar-approx}) in Eq.~(\ref{conf-scalar-eqs-1}), we get
\begin{align}
\notag
\widehat{F}^{IJ}_{\alpha \mu} e^\alpha_I e_{\nu J} - \frac12 g_{\mu\nu} \widehat{F}^{IJ}_{\alpha \beta} e^\alpha_I e^\beta_J = \kappa \left( \partial_\mu \phi \partial_\nu \phi - \frac12 g_{\mu\nu} g^{\alpha\beta} \partial_\alpha \phi \partial_\beta \phi + \frac16  (\widehat{F}^{IJ}_{\alpha \mu} e^\alpha_I e_{\nu J} - \frac12 g_{\mu\nu} \widehat{F}^{IJ}_{\alpha \beta} e^\alpha_I e^\beta_J)\phi^2 \right.\\
\left. + \frac{1}{6} \left[ g_{\mu\nu} \widehat{\nabla}_\sigma \widehat{\nabla}^\sigma\phi^2 
- \widehat{\nabla}_\mu \widehat{\nabla}_\nu \phi^2 \right] \right) + \mathcal{O}(\kappa^2)\,.
\end{align}
We have thus obtained the correct energy-momentum tensor, with $ \mathcal{O}(\kappa^2) $ 
contributions which can be neglected. There is however another problem with Eq.~(\ref{conf-scalar-eqs-3}). Let us insert Eq.~(\ref{spin-con-scalar-approx}) in Eq.~(\ref{conf-scalar-eqs-3}) to write it up to $\mathcal{O}(\kappa)\,$ as
\begin{align}\label{conf-scalar-eqs-3-torsion}
	\widehat{\nabla}_\mu \widehat{\nabla}^\mu \phi - \frac16 \widehat{F}^{IJ}_{\alpha \beta} e^\alpha_I e^\beta_J \phi + \frac{\kappa}{24} \phi \widehat{\nabla}_\alpha \widehat{\nabla}^\alpha \phi^2 = 0\,.
\end{align}
Upon a conformal transformation this equation becomes
\begin{align}
	\Omega^{-3}\left(\widehat{\nabla}_\mu \widehat{\nabla}^\mu \phi - \frac16 \widehat{R} \phi \right) + \frac{\kappa}{24}\Omega^{-5} \phi \widehat{\nabla}_\alpha \widehat{\nabla}^\alpha \phi^2 -\frac{\kappa}{6} \Omega^{-5} \phi^2 g^{\alpha\beta} \partial_\alpha\phi \partial_\beta\phi - \frac{\kappa}{12} \Omega^{-5} \phi^3 \widehat{\nabla}_\alpha \widehat{\nabla}^\alpha \ln \Omega = 0\,.
\end{align}
Clearly the equation is not conformally covariant. The last term in Eq.~(\ref{conf-scalar-eqs-3-torsion}) not only has different conformal weight, it also transforms inhomogeneously. The source of this discrepancy lies in the expression of spin connection, more precisely the contorsion $ \Lambda\,, $ which transforms as
\begin{align}
\Lambda^{IJ}_\mu \rightarrow \Omega^{-2}\Lambda^{IJ}_\mu + \frac{\kappa}{6} \Omega^{-2} \left(e^I_\mu e^{J\alpha} - e^J_\mu e^{I\alpha}\right) \phi^2 \partial_\alpha \ln\Omega\,.
\end{align}
The conformal weight of $\Lambda^{IJ}_\mu$ is different from that of the torsion-free part $ \omega^{IJ}_\mu $ of Eq.~(\ref{ax}) and also it transforms inhomogeneously. 
We can resolve this problem by modifying the action such that it does not produce torsion dynamically.
The connection used in the VEP formalism is 
not torsion-free a priori. However, the
$R\phi^2$, in the metric formalism, is constructed from a torsion-free connection.
Therefore, in the VEP formalism if we construct the non-minimal 
$R\phi^2$ term only from the torsion-less part of the connection we can eliminate on-shell
torsion.
The total action now reads 
\begin{align}\label{scalar-action-VEP-mod}
S[e,A,\phi] = \frac{1}{2\kappa} \int |e|d^4x F^{IJ}_{\mu\nu}e^\mu_I e^\nu_J   + \int |e|d^4x \left( -\frac12 e^\mu_I e^{\nu I} \partial_\mu \phi \partial_\nu \phi - \frac{1}{12}\widehat{F}[\omega]^{IJ}_{\mu\nu}e^\mu_I e^\nu_J\phi^2\right)\,.
\end{align}
The equations of motion as obtained from this action are 
\begin{subequations}\label{conf-scalar-eqs-mod}
	\begin{align}
	\notag
	\delta e^\nu_J &: R_{\mu\nu} - \frac12 g_{\mu\nu} R = \kappa \left( \partial_\mu \phi \partial_\nu \phi - \frac12 g_{\mu\nu} g^{\alpha\beta} \partial_\alpha \phi \partial_\beta \phi + \frac16  (\widehat{R}_{\mu\nu} - \frac12 g_{\mu\nu} \widehat{R})\phi^2 \right.\\
	&\qquad \qquad \qquad \qquad \qquad \left. + \frac{1}{6} \left[ g_{\mu\nu} \widehat{\nabla}_\sigma \widehat{\nabla}^\sigma\phi^2 
	- \widehat{\nabla}_\mu \widehat{\nabla}_\nu \phi^2 \right] \right) , \label{conf-scalar-eqs-mod-1}\\
	\delta A^{IJ}_\nu &: A^{IJ}_\mu = \omega^{IJ}_\mu [e] \label{conf-scalar-eqs-mod-2},\\
	\delta \phi &: \widehat{\nabla}_\mu \widehat{\nabla}^\mu \phi - \frac16 \widehat{R} \phi=0 \label{conf-scalar-eqs-mod-3}\,.
	\end{align}
\end{subequations}
The scalar field equation~(\ref{conf-scalar-eqs-mod-3}), thus gets back the torsion-free, conformally invariant form given by Eq.~(\ref{conf-scalar-eqn}). The action in Eq.~(\ref{scalar-action-VEP-mod}) has enabled us to eliminate on-shell torsion and identify the equations with those in the usual metric formalism. Also upon using the on-shell expression of the spin connection given by Eq.~(\ref{conf-scalar-eqs-mod-2}), the left hand side of Einstein's equation~(\ref{conf-scalar-eqs-mod-1}) becomes torsion-free and the equation can be identified with that in the tetrad formalism given in Eq.~(\ref{conf-scalar-eqs-tetrad}).

\subsection{Dynamically generated torsion and fermion}\label{fermion.torsion}
Let us now consider the conformal transformation of fermionic field equation when the on-shell torsion 
arises from the field itself. The expression for the on-shell torsion with fermionic field was found in Eq.~(\ref{torsion}), and using this expression, we got a non-linear Dirac equation
\begin{align}
\gamma^K e^\mu_K  \partial_\mu \psi - \frac{i}{4}\omega^{IJ}_\mu e^{\mu K} \gamma_K \sigma_{IJ} \psi -\frac{i\kappa}{64}\bar{\psi}\{\gamma_K,\sigma_{IJ}\}\psi\{\gamma^K,\sigma^{IJ}\}\psi  = 0\,.
\end{align}
Comparing with the Dirac equation in the tetrad formulation, we recognize that the first two terms are covariant under 
conformal transformations. The cubic term however transforms with a different weight,
as can be seen from the transformed equation,
\begin{align}
\Omega^{-\frac52}\left(\gamma^K e^\mu_K  \partial_\mu \psi - \frac{i}{4}\omega^{IJ}_\mu e^{\mu K} \gamma_K \sigma_{IJ} \psi\right) -\frac{i\kappa}{64}\Omega^{-\frac92}\bar{\psi}\{\gamma_K,\sigma_{IJ}\}\psi\{\gamma^K,\sigma^{IJ}\}\psi = 0\,.
\end{align}
The nonlinear Dirac equation is thus not invariant under conformal transformation.
The source of this difference in conformal weight is as follows.

The cubic term which breaks the invariance of the equation appeared due to the dynamically generated contorsion $\Lambda$\,.
Let us consider the transformation of the spin connection of Eq.~(\ref{complete-action-eqs-2}) to see this.
\begin{align}\label{spin-con-trans-spinor}
A^{IJ}_\mu \quad \rightarrow \quad \omega^{IJ}_\mu [e] + \left( e^I_\mu e^{J\nu} - e^J_\mu e^{I\nu} \right) \partial_\nu \ln \Omega +  \Omega^{-2} \Lambda^{IJ}_\mu\,.
\end{align}
The two components of the spin connection $ \omega $ and $ \Lambda $ transform with different conformal weights. 
This is because they come from two different sectors of the theory. While $ \omega $ comes from the gravity sector
and is fully determined by the geometry, $ \Lambda $ comes from the matter sector, in this case fermions. 
But these two sectors have different conformal weights. 
The fermion Lagrangian has weight $ -4 $ but $ R $ in the gravity part has weight of $ -2 $. This difference shows 
up in the transformation of $ \Lambda $. The transformation of the spin connection also implies that torsion 
as given by Eq.~(\ref{torsion}) transforms homogeneously with conformal weight $-2$\,,
\begin{equation}
\label{torsion_conf}
~^{OS}C^\nu_{\phantom{\nu}\rho\lambda} \rightarrow \Omega^{-2} ~^{OS}C^\nu_{\phantom{\nu}\rho\lambda}\,.
\end{equation}
It is thus clear that on-shell torsion tensor (with index positions as above) transforms homogeneously unlike in Nieh-Yan theory where it has an inhomogeneous transformation. The above transformation is similar to that of the invariant torsion of Eq.~(\ref{Inv-torsion-conf}) apart from an overall factor of $\Omega^{-2}\,$.
We can thus conclude that similar to what has been observed in the case of scalar field, dynamically generated torsion also breaks the conformal invariance of Dirac equation. We can try to make the Dirac equation conformally invariant but this requires the fundamental theories to be modified as we will discuss below.

There are different ways to proceed if we wish to reinstate conformal invariance.
First, if we have an action for gravity that scales in the same manner as the matter action, we can eliminate the above mentioned weight difference of $\omega$ and $\Lambda$ altogether. A scale invariant theory of gravity with spinors, as discussed in~\cite{Dereli:1982xb} where the authors considered Brans-Dicke (BD) theory of gravity, might be able to provide a way out. If it is possible to construct such a theory of gravity without affecting the on-shell quantities obtained in this paper, we will get back the conformally covariant Dirac equation. 
Another way was discussed in~\cite{Fabbri:2011ha} where an invariant theory of Dirac field was considered in conformal gravity such that the non-linearities in the Dirac equation do not appear.

A third way of making the Dirac equation conformally invariant is to modify the fermionic Lagrangian. If we are interested in the conformal invariance of the Dirac equation alone, we can modify the spinor Lagrangian by the addition of a quartic term. We choose the term it in such a way that it cancels the cubic term thereby making the equation linear. We should also keep in mind that this term should not affect the expression for on-shell torsion i.e., the spin connection should not appear in it. We modify the spinor Lagrangian as
\begin{align}
\notag
\mathcal{L}_{F-mod} = \frac{i}{2}	\left(\bar{\psi}\gamma^K e^\mu_K\partial_\mu\psi - \partial_\mu\bar{\psi}\gamma^K e^\mu_K\psi - \frac{i}{4} A^{IJ}_\mu e^{\mu K} \bar{\psi}\{\gamma_K,\sigma_{IJ} \} \psi\right.\\
\left. +\frac{i\kappa}{64}\bar{\psi}\{\gamma_K,\sigma_{IJ}\}\psi\bar{\psi}\{\gamma^K,\sigma^{IJ}\}\psi\right)\,.
\end{align}
The equation obtained by extremising the corresponding action with respect to $\bar{\psi}$ is
\begin{align}
\gamma^K e^\mu_K  \partial_\mu \psi - \frac{i}{4}A^{IJ}_\mu e^{\mu K} \gamma_K \sigma_{IJ} \psi +\frac{i\kappa}{64}\bar{\psi}\{\gamma_K,\sigma_{IJ}\}\psi\{\gamma^K,\sigma^{IJ}\}\psi  = 0
\end{align}
Also, upon extremisation with the spin connection, we get exactly the same expressions for the contorsion 
$\Lambda$ and torsion $C$ as found in Eq.~(\ref{lambda-vep}) and Eq.~(\ref{torsion}) earlier. So when we 
put the on-shell expression of the spin connection in the above equation we get nothing but the linear 
Dirac equation which we had obtained in tetrad formulation i.~e.,
\begin{align}
\gamma^K e^\mu_K  \partial_\mu \psi - \frac{i}{4}\omega^{IJ}_\mu e^{\mu K} \gamma_K \sigma_{IJ} \psi = 0\,.
\end{align} 
We have thus eliminated the cubic term and obtained the conformally invariant Dirac equation. The added term, however, also cancels the $\mathcal{O}(\kappa^2)$ on the right hand side of Einstein's equations~(\ref{EE-spinor-sym}). In other words, with the addition of the term we are effectively dealing with a torsion-free theory.

Also, unlike in Nieh-Yan theory (and invariant torsion), the Lagrangian above is not conformally invariant although the equation it produces is invariant under conformal transformation as we have seen above. 
%

\section{Discussion}\label{conclusion}
We have discussed two possibilities of the transformation of the spin connection $A^{IJ}_\mu$ in the level of action. The two cases can be parametrised with a single parameter. In order to do that, we recall that the torsion-free part of the connection can be written as
\begin{align}
	\omega^{IJ}_\mu = A^{IJ}_\mu - \Lambda^{IJ}_\mu\,.
\end{align} 
The transformation of $\omega $ is dictated from that of the tetrads and this implies that the right hand side of the equation has a definite conformal transformation. If we consider the case where $A^{IJ}_\mu$ remains invariant, $\Lambda^{IJ}_\mu$ must transform like $\omega^{IJ}_\mu$ with a negative sign (Nieh-Yan). On the other hand, if it is $\Lambda^{IJ}_\mu$ that remains invariant, $A^{IJ}_\mu$ must transform like $\omega^{IJ}_\mu$ (Invariant torsion). The following transformations
\begin{align}
&A^{IJ}_\mu \rightarrow A^{IJ}_\mu + \xi \left(e^I_\mu e^{J\nu} - e^J_\mu e^{I\nu}\right)\partial_\nu\ln\Omega\\
&\Lambda^{IJ}_\mu \rightarrow \Lambda^{IJ}_\mu - (1-\xi) \left(e^I_\mu e^{J\nu} - e^J_\mu e^{I\nu}\right)\partial_\nu\ln\Omega\,,
\end{align}
interpolate between the two cases with $\xi=0$ and $\xi=1\,,$ with Nieh-Yan theory
corresponding to $\xi=0$ and invariant torsion corresponding to $\xi=1$. 
Other values 
of $\xi$ also correspond to conformal transformations, with the Dirac
Eq.~(\ref{Dirac-eq-wo-skew}) remaining invariant for any value of $\xi$ between $0$ and $1$.
This parametrization may be compared with that postulated for the transformation of torsion in conformal gravity~\cite{Fabbri:2011vk}, where it is an undetermined constant.
For the scalar field however, the non-minimal
coupling term $-\frac{1}{12}R\phi^2$ must be written using the torsion-free Ricci scalar in order for the field equation to be conformally invariant, as was observed for both Nieh-Yan theory and invariant torsion.

We see that dynamically generated torsion does not transform inhomogeneously at all. There is of course an issue with torsion if considered to be given by the equations of motion alone: the contorsion part has a relative weight over the torsion-free part.

For minimally coupled fermionic fields, the on-shell torsion resulting from the fermion 
coupling makes the Dirac equation non-linear. This equation is not conformally invariant.  
We have seen that there are different ways of restoring conformal invariance of 
the Dirac equation, one of which involves the addition of a quartic term in the Lagrangian. 
This term helps us recover the conformally invariant linear Dirac equation by setting the 
torsion to vanish on shell, but the cost is the additional quartic term. We note that it 
is also possible to consider the torsion, or alternatively the fermion, as a source of 
explicit breaking of conformal symmetry, as was done in~\cite{Fabbri:2012ws}.

\end{document}